\documentclass[prl,reprint, twocolumn,showpacs,preprintnumbers,amsmath,amssymb,nofootinbib,floatfix]{revtex4-1} 
\usepackage{graphicx}

\usepackage{mathrsfs}
\usepackage{hyperref}

\usepackage{slashed}

\usepackage{color}

\usepackage{gensymb}

\usepackage{amsmath}
\allowdisplaybreaks[4]

\def \matrix #1 {\left(\begin{array}{cc} #1 \end{array}\right)}

\def\II{\hbox{{1}\kern-.25em\hbox{l}}}

\newcommand{\Mahogany}[1]{\textcolor[rgb]{0.75,0.25,0.00}{#1}}

\begin{document}

\title{Next-to-Leading-Order QCD Predictions for the Nucleon Form Factors }

\author{Yong-Kang Huang}
\email{huangyongkang@mail.nankai.edu.cn}

\author{Bo-Xuan Shi}
\email{corresponding author: shibx@mail.nankai.edu.cn}

\author{Yu-Ming Wang}
\email{corresponding author: wangyuming@nankai.edu.cn}

\author{Xue-Chen Zhao}
\email{corresponding author: zxc@mail.nankai.edu.cn}

\affiliation{\vspace{0.2 cm}
School of Physics, Nankai University, \\
Weijin Road 94, Tianjin 300071, P.R. China \\}

\date{\today}

\begin{abstract}
\noindent
We accomplish for the first time the next-to-leading-order QCD computations of the leading-twist contributions
to the Dirac form factors of  both the proton and the neutron by applying the hard-collinear factorization theorem rigorously.
The resulting predictions for these baryon form factors indicate that the one-loop perturbative corrections
to the hard-gluon-exchange contributions are numerically substantial for a wide range of  momentum transfers
accessible  in  the current and forthcoming collider experiments.
Including further the (formally) power-suppressed soft contributions due to the celebrated Feynman mechanism,
we then perform the state-of-the-art analysis of the Dirac electromagnetic nucleon form factors
from first field-theoretical principles, thus allowing for the most robust determinations of the nucleon distribution amplitudes
from the direct comparison with the experimental measurements.
\\[0.4em]

\end{abstract}


\maketitle

%
\section{Introduction}
%

It is generally accepted that the nucleon electromagnetic form factors are
among the simplest, most fundamental, and best-known  measurable quantities
for exploring diverse facets of the non-perturbative QCD dynamics
that govern  the complex nature  of  nucleon structure properties
and for  advancing our understanding towards the perturbative factorization formalism
that enable us to separate the short-distance QCD interaction from
the long-range strong interaction dynamics for the entire hard exclusive processes
with sufficiently large momentum transfers.
The pioneering investigation of nucleon tomography dates back to
the elastic electron-nucleon scattering experiment in the early nineteen fifties,
long before the advent of QCD,  conducted by Robert Hofstadter and collaborators  \cite{Hofstadter:1953zjy,Hofstadter:1955ae,Yearian:1958shh,Hofstadter:1956qs}
in the High Energy Physics Laboratory (HEPL) at Stanford University.
From then on, probing  the internal structure of the composite nucleon state
with unprecedented precision has become the central focus of the world-wide hadron-physics programs,
reflected by the increasing number of electron accelerators and laboratories over the past decades
(see for instance \cite{Price:1971zk,Hanson:1973vf,Berger:1971kr,Bartel:1973rf,Bosted:1992rq,Andivahis:1994rq,
JeffersonLabHallA:1999epl,ResonanceSpinStructure:2006oim,Arrington:2021alx,Accardi:2012qut,Schmookler:2022gxw,Accardi:2023chb}).
Extensive experimental efforts in measuring the nucleon electromagnetic form factors
with  the Rosenbluth cross section separation method and with the polarization transfer technique
have been well documented in several review articles
(see  \cite{Gao:2003ag,Hyde:2004gef,Arrington:2006zm,Pacetti:2014jai,Gross:2022hyw} for an incomplete list).

Needless to say,   the ever-lasting and tremendous experimental investment in exploiting the emergent nucleon structure
naturally triggers  a considerable amount of theoretical developments to achieve  the  field-theoretic description
of hard exclusive reactions, thus resulting in a variety of technical frameworks and phenomenological methods
(see \cite{Coriano:1998ge,Perdrisat:2006hj,Pire:2021hbl,Gross:2022hyw,Stefanis:1997zyh,Aznauryan:2012ba} for an overview).
It has been demonstrated that  the underlying QCD dynamics of the nucleon electromagnetic form factors is
dictated by the delicate interplay of two competing mechanisms (namely, the soft  Feynman mechanism versus
the hard-scattering mechanism).
A systematic and successful approach to address the soft non-factorizable contributions to the nucleon transition form factors
has been formulated in the seminal works  \cite{Braun:2001tj,Braun:2006hz,Braun:2005be},
by adopting the  light-cone sum rules (LCSR) technique \cite{Balitsky:1989ry,Chernyak:1990ag},
which receives numerous applications in the QCD-based calculations of  $B$-meson decay matrix elements
\cite{Ball:1998kk,Ball:2004ye,Duplancic:2008ix,Khodjamirian:2011ub,Khodjamirian:2010vf,Khodjamirian:2012rm,Khodjamirian:2023wol}
and of heavy-baryon decay form factors \cite{Wang:2008sm,Wang:2009hra,Khodjamirian:2011jp,Feldmann:2011xf,Wang:2015ndk,Huang:2022lfr,Feldmann:2023plv}.
Subsequently,  the next-to-leading order (NLO) QCD  corrections  to the LCSR  predictions for the nucleon form factors
have been computed at the twist-three accuracy \cite {Passek-Kumericki:2008uqr}  with  the covariant trace formalism
and at the twist-four accuracy \cite{Anikin:2013aka} with the Krankl-Manashov (KM) renormalization scheme
for the three-body light-ray operators \cite{Krankl:2011gch}.
On the other hand, the hard-gluon-exchange contributions to these nucleon form factors can be further computed
with the same LCSR framework by carrying out two-loop computations of the considered correlation functions,
at the price of introducing  slight model dependence of the nucleon disentanglement from the higher-mass background \cite{Braun:1997kw,Ball:1997rj}.

Instead, an elegant and  model-independent description of  the hard-scattering contributions
to the nucleon electromagnetic form factors  can be routinely  constructed with the perturbative factorization formalism,
based upon the diagrammatic analysis \cite{Lepage:1980fj,Efremov:1979qk,Chernyak:1983ej} and the effective field theory approach \cite{Kivel:2010ns,Kivel:2012mf}.
As a matter of fact, the textbook tree-level computation of the nucleon form factors
has been accomplished with the hard-collinear factorization prescription more than forty years ago
\cite{Lepage:1979za,Chernyak:1984bm} (see \cite{Ji:1986uh,Carlson:1987sw,Stefanis:1987vr, Brooks:2000nb,Thomson:2006ny}
for  subsequent  re-computations and \cite{Isgur:1988iw,Radyushkin:1998rt,Bolz:1996sw,Brodsky:1989pv,Botts:1989kf,Li:1992nu}
for additional  discussions on the applicability of perturbative QCD at accessible energies).
Apparently,  extending the leading order (LO) QCD calculation of the  hard-scattering contributions
to the NLO accuracy (see \cite{Knodlseder:2015vmu} for an earlier attempt)
will therefore be in high demand for validating explicitly the collinear factorization scheme
of baryon transition matrix elements truly at the quantum level and for unraveling the intriguing pattern of perturbative expansion
for multi-parton scattering amplitudes.
In this Letter, we will employ the modern QCD factorization formalism to extract the NLO short-distance coefficient function
analytically, by evaluating the desired $7$-point partonic amplitude at ${\cal O}(\alpha_s^3)$
with the nowadays standard one-loop computational prescriptions
and by performing the ultraviolet (UV) renormalization  and  infrared (IR) subtractions
in a rigorous and  factorization-compatible fashion.
An emphasis will be then placed on  phenomenological impacts of the thus determined radiative corrections
to the hard-gluon-exchange contributions
in predicting the nucleon form factors at large momentum transfers,
by taking advantage of three sample models of the leading-twist nucleon distribution amplitude.

%
\section{Electromagnetic Nucleon Form Factors}
%

We first set up the theory framework for constructing the  QCD factorization formulae of
the nucleon  form factors   at large momentum transfers.
Adopting the customary definitions of the electromagnetic form factors of the nucleon
enables us to write down \cite{Foldy:1952zz}
\begin{eqnarray}
\langle N(P^{\prime})  | j_{\mu}^{\rm em}(0) | N(P) \rangle
&=&   \bar N(P^{\prime}) \,  \big [ \gamma_{\mu} \, F_1 (Q^2)
\nonumber \\
&&  - \, i \, \frac{\sigma_{\mu \nu} \, q^{\nu}}{m_N}  \, F_2 (Q^2) \big  ]  \, N(P) \,,
\label{definition of nucleon form factors}
\end{eqnarray}
where $P$  ($P^{\prime}$) stands for  the four-momentum carried by the initial (final) nucleon state,
$q=P-P^{\prime}$ refers to the transfer momentum,
and $N(P)$ ($N(P^{\prime})$) corresponds to the on-shell nucleon spinor.
In addition, we have employed  the convention $\sigma_{\mu \nu} = (i /2) \, [\gamma_{\mu}, \, \gamma_{\nu}]$
and the notation $Q^2=-q^2$.
The electromagnetic current for the active quark is given by
\begin{eqnarray}
j_{\mu}^{\rm em}(x)  =  Q_{u} \, \bar u(x) \gamma_{\mu} u(x) +  Q_{d} \, \bar d(x) \gamma_{\mu} d(x)\,.
\label{definition of the EM current}
\end{eqnarray}
Introducing further two light-cone vectors $n_{\mu}$ and $\bar n_{\mu}$
with  the constraints $n^2=\bar n^2=0$ and $n \cdot \bar n=2$
leads to the decomposition $P_{\mu} \approx  ({n \cdot P} /2) \, \bar n_{\mu}$
and $P^{\prime}_{\mu} \approx ({\bar n \cdot P^{\prime}} /2) \, n_{\mu}$
at leading power in $\Lambda_{\rm QCD}^2/Q^2$,
where $\Lambda_{\rm QCD}$ denotes the scale of the strong interaction.
The Dirac  form factor $F_1(Q^2)$ characterizes  the charge and ``non-anomalous" magnetic moment distribution  in the nucleon,
while the Pauli form factor $F_2(Q^2)$ measures the anomalous magnetic moment distribution in the nucleon
(see \cite{Perdrisat:2006hj,Gross:2022hyw} for an overview).
Importantly,  the hard-scattering contribution to the Dirac form factor exhibits the scaling behaviour of $(\Lambda_{\rm QCD}/Q)^4$
(modulo logarithms of $Q^2$) in the asymptotic limit $Q^2 \to \infty$ \cite{Brodsky:1973kr,Matveev:1973ra,Belitsky:2002kj}.
By contrast, the asymptotic prediction of  the helicity-flip Pauli form factor turns out to be suppressed
by an extra power of $\Lambda_{\rm QCD}^2/Q^2$.
As a consequence, we will  concentrate on the NLO QCD computation of the hard-gluon-exchange contribution to
the  Dirac  nucleon  form factor $F_1 (Q^2)$.

According to the hard-collinear factorization prescription  \cite{Lepage:1980fj,Chernyak:1983ej},
the leading-power contribution  to the Dirac  form factor  in the large-momentum expansion
can be cast  in the form of (see \cite{Duncan:1979hi,Duncan:1979ny,Mueller:1981sg,Milshtein:1981cy,Milshtein:1982js,Kivel:2010ns,Kivel:2012mf}
for potential complications beyond the ${\cal O}(\alpha_s^3)$ approximation)
\begin{eqnarray}
F_1(Q^2) &=& \frac{(4  \pi  \alpha_s)^2} {Q^4}  \,  \int [{\cal D} x] \int [{\cal D}  y ] \,\,
 {\mathbb T}_{N}(x_i,  y_i,  Q^2, \mu_F)
\nonumber \\
&&  \times \, \varphi_N(x_i, \mu_F)  \,\,   \varphi_N(y_i, \mu_F)  \,,
\label{hard-collinear factorization formula}
\end{eqnarray}
where the integration measure is defined as
\begin{eqnarray}
\int [{\cal D} z]  = \int_0^1 \, d z_1 \, d z_2 \, d z_3  \,\,   \delta(z_1 + z_2 + z_3 - 1) \,,
\end{eqnarray}
and  $\mu_F$ refers to the factorization scale corresponding to
the resolution with which the  nucleon structure is being probed.
The short-distance coefficient function ${\mathbb T}_{N}$ can be expanded perturbatively
in terms of the renormalized coupling constant  (similarly for any other QCD quantity)
\begin{eqnarray}
{\mathbb T}_{N} = {\mathbb T}_{N}^{(0)} +  \left ({\alpha_s \over 4 \, \pi} \right ) \, {\mathbb T}_{N}^{(1)}  + {\cal O}(\alpha_s^2) \,.
\label{expansion of the final hard kernel}
\end{eqnarray}
The twist-three nucleon distribution amplitude $\varphi_N$ can be defined by the renormalized matrix element
of the leading three-body light-ray   operator  \cite{Chernyak:1983ej,Braun:2000kw}
\begin{widetext}
\begin{eqnarray}
&& \left \langle 0 \left |   \epsilon_{i j k} \, \left [  u_{i^{\prime}}^{\uparrow}(\tau_1 \, n)[\tau_1 \, n,  \, \tau_0 \, n]_{i^{\prime} i} \,\, C \,  \slashed{n} \,\,  u_{j^{\prime}}^{\downarrow}(\tau_2 \, n)[\tau_2 \, n,  \, \tau_0 \, n]_{j^{\prime} j}  \right ] \,
\slashed{n} \,  d_{k^{\prime}}^{\uparrow}(\tau_3 \, n)[\tau_3 \, n,  \, \tau_0 \, n]_{k^{\prime} k}
\right | N^{\uparrow} (P) \right  \rangle
\nonumber \\
&& = - {f_N(\mu_F) \over 2} \, (n \cdot P)  \, \slashed{n} \,N^{\uparrow}(P) \,
\int [{\cal D} x] \, {\rm exp} \left [ - \, i \, n \cdot P \,  \sum_{i=1}^{3} \, x_i \, \tau_i \right ] \, \varphi_N(x_i, \mu_F)  \,,
\label{definiton of the twist-3 nucleon DA}
\end{eqnarray}
\end{widetext}
where $[\tau_i \, n,  \, \tau_0 \, n]$ is the collinear Wilson line to ensure gauge invariance \cite{Braun:1999te}
and $C$ is the charge conjugation matrix.
We have also introduced the   shorthand notation for the quark fields with definite chirality
$q^{\uparrow (\downarrow)} = {1 \over 2} \, (1 \pm \gamma_5)  \, q$.
The renormalization-group (RG)   equation of the QCD string operator
on the left-hand side of (\ref{definiton of the twist-3 nucleon DA}) \cite{Balitsky:1987bk}
motivates the conformal expansion of $\varphi_N$ over a set of orthogonal polynomials
 defined as eigenfunctions of the corresponding one-loop evolution kernel
\begin{eqnarray}
\varphi_N(x_i, \mu_F)  = 120 \, x_1 x_2 x_3  \, \sum_{n=0}^{\infty} \, \sum_{k=0}^{n} \,
\varphi_{n k}(\mu_F)  \, {\cal P}_{nk}(x_i) \,,
\hspace{0.6 cm}
\end{eqnarray}
where the polynomials ${\cal P}_{nk}$ have definite parity under the permutation of $x_1 \leftrightarrow x_3$
and their manifest expressions  with  $n = 0, 1, 2$ have been  displayed in \cite{Braun:2008ia}.
We employ the normalization condition $ \int [{\cal D} x] \, \varphi_N(x_i, \mu_F)=1$
such that $\varphi_{0 0} = 1$.
The normalization constant $f_N$ and the shape parameters $\varphi_{n k}$ ($n\geq 1$) are defined by
non-perturbative matrix elements of  local gauge-invariant three-body operators \cite{Braun:2014wpa}.

The short-distance matching coefficient ${\mathbb T}$ can be conveniently determined by
 inspecting the $7$-point QCD matrix element
\begin{eqnarray}
\Pi_{\mu} =  \langle u(P_1^{\prime}) \, u(P_2^{\prime}) \, d(P_3^{\prime})
| j_{\mu}^{\rm em}(0) |  u(P_1) \, u(P_2) \, d(P_3) \rangle \,,
\hspace{0.5 cm}
\end{eqnarray}
where the external parton momenta  can be restricted to their leading components
$P_i = x_i \, P$ and $P_i^{\prime} = y_i \, P^{\prime}$ at the leading power accuracy.
We will employ  dimensional regularization with $D = 4 - 2 \, \epsilon$ in the entire calculation
of the bare one-loop amplitude.
Apparently, the partonic quantity $\Pi_{\mu}$ contains  both UV and IR (collinear and anti-collinear) singularities.
The former divergences are cancelled by the UV renormalization for the light quark fields in the on-shell scheme
and for the strong coupling $\alpha_s$ in the ${\rm \overline{MS}}$ scheme \cite{Beneke:2008ei};
the latter are subtracted by the nucleon distribution amplitude contributions
with the aid of the  prescription for an appropriate treatment of  evanescent operators \cite{Dugan:1990df,Herrlich:1994kh,Wang:2017ijn,Gao:2021iqq}.

%
\section{Next-to-Leading-Order QCD Computations}
%

\begin{figure}[htp]
\includegraphics[width=0.45 \textwidth]{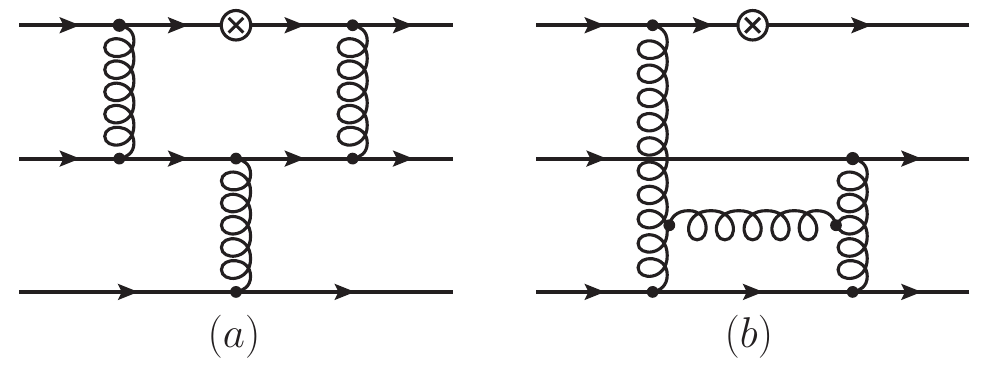}
\caption{Sample Feynman diagrams at NLO, where the circled cross marks an insertion of the electromagnetic current. }
\label{fig:1-loop-diagrams}
\end{figure}

In this section we will describe briefly the one-loop QCD computation of the bare amplitude $\Pi_{\mu}$.
We   generate the NLO Feynman diagrams with {\tt FeynArts} \cite{Hahn:2000kx}
and simultaneously by means of an in-house Mathematica routine.
Taking into account the observation that  certain Feynman diagrams cannot contribute
due to the QCD equations of motion, the vanishing colour (and/or electric charge) factors,
as well as the Furry  theorem, we obtain $1371$ non-vanishing diagrams at one loop eventually.
The sample Feynman diagrams  are explicitly shown in Figure \ref{fig:1-loop-diagrams}.

The Passarino-Veltman  reduction \cite{Passarino:1978jh} to the tensor structure of the amplitude is implemented first,
and the Dirac and colour algebras are then performed with in-house routines.
The yielding  one-loop scalar integrals  are further reduced to a small set of master integrals
with   the packages  {\tt Apart} \cite{Feng:2012iq} and   {\tt FIRE} \cite{Smirnov:2008iw}.
Exploiting further the fact that  $P_{i}  \, \| \, P $ and  $P_i^{\prime}  \, \| \, P^{\prime}$,
we   arrive at {\it only two} linearly independent master integrals at one loop
(with ${\cal Q}_{1, 2}^2 \neq 0$, $({\cal Q}_1-{\cal Q}_2)^2 \neq 0$ and  ${\cal Q}^2 \neq 0$)
\begin{eqnarray}
{\cal I}_{\rm A} &=&  \int {d^D \ell \over (2 \pi)^D} \, \frac{1}{[\ell^2 + i 0] }
\, \frac{1}{[(\ell + {\cal Q}_1)^2 + i 0] } \, \frac{1}{[(\ell + {\cal Q}_2)^2 + i 0] } \,,
\nonumber \\
{\cal I}_{\rm B} &=&  \int {d^D \ell \over (2 \pi)^D} \, \frac{1}{[\ell^2 + i 0] }
\, \frac{1}{[(\ell + {\cal Q})^2 + i 0] }  \,,
\end{eqnarray}
which can be readily expressed in terms of polylogarithms and logarithms in the analytic $\epsilon$ expansion.
It is straightforward to verify that the one-loop $3$-point integral ${\cal I}_{\rm A}$ for ${\cal Q}_{1 (2)}^2 = 0$
can be obtained as a linear combination of ${\cal I}_{\rm B}$.

%
\section{Hard-Collinear Factorization Formula}
%

We will dedicate this section to the derivation of the master formula for the hard-scattering kernel
by performing the UV renormalization and IR subtractions.
This can be routinely achieved by matching the QCD amplitude  $\Pi^{\mu}$ onto the collinear matrix elements
\begin{eqnarray}
\Pi^{\mu} &=& \frac{(4 \pi \alpha_s)^2}{Q^6} \, \sum_{k}  \,
T_{k} \otimes \langle {\cal O}_{k}^{\mu} \rangle  \,,
\nonumber \\
{\cal O}_{k}^{\mu} & \in &   \left \{  {\cal O}_{N}^{\mu}, \,  {\cal O}_{T}^{\mu}, \,
E_{1}^{\mu}, \, E_{2}^{\mu}, \, E_{3}^{\mu}  \right \} \,,
\label{matching condition}
\end{eqnarray}
where we have to extend the effective operator basis by including the so-called evanescent operators
\begin{eqnarray}
{\cal O}_{N}^{\mu}  &=&    {\color{blue} [ \bar \chi_{u}^{\downarrow}  \slashed{\bar n}  C^{-1}
\bar \chi_{u}^{\uparrow}  ]  \,
 [ \bar \chi_{d}^{\uparrow}  \slashed{\bar n} } \,  \gamma_{\perp}^{\mu} \,
\Mahogany { \slashed{n}   \xi_{d}^{\uparrow}  ] \,
 [  \xi_{u}^{\uparrow}  C \slashed{n}  \xi_{u}^{\downarrow} ]},
\nonumber \\
{\cal O}_{T}^{\mu}  &=&  {\color{blue}  [ \bar \chi_{u}^{\uparrow}  \slashed{\bar n}  \gamma_{\perp \beta}  C^{-1}
\bar \chi_{u}^{\uparrow}  ] \,
[ \bar \chi_{d}^{\downarrow} \, \slashed{\bar n}   \gamma_{\perp}^{\beta}}  \, \gamma_{\perp}^{\mu}  \,
\Mahogany {  \gamma_{\perp}^{\alpha}  \slashed{n}  \xi_{d}^{\downarrow} ] \,
[ \xi_{u}^{\uparrow} C  \gamma_{\perp \alpha} \slashed{n} \xi_{u}^{\uparrow} ]},
\nonumber \\
E_{1}^{\mu}  &=& [\bar \chi_u \gamma_{\perp}^{\mu} \xi_u  ] \,
 [\bar \chi_u \gamma_{\perp}^{\alpha} \xi_u ] \,
[\bar \chi_d \gamma_{\perp \alpha} \xi_d ]
 -  {1 \over 8} \, {\cal O}_{N}^{\mu} -  {1 \over 2} \, {\cal O}_{T}^{\mu},
\nonumber \\
E_{2}^{\mu}  &=&  [\bar \chi_u \gamma_{\perp}^{\alpha} \gamma_{\perp}^{\beta} \gamma_{\perp}^{\mu} \xi_u  ] \,
 [\bar \chi_u \gamma_{\perp \beta}  \xi_u ] \,
[\bar \chi_d \gamma_{\perp \alpha}  \xi_d ]   -  {\cal O}_{T}^{\mu},
\nonumber \\
E_{3}^{\mu}  &=&   [\bar \chi_u \gamma_{\perp \alpha}  \xi_u  ] \,
 [\bar \chi_u \gamma_{\perp \beta}  \xi_u ] \,
[\bar \chi_d \gamma_{\perp}^{\beta}  \gamma_{\perp}^{\alpha} \gamma_{\perp}^{\mu}  \xi_d ]    - {1 \over 4} \, {\cal O}_{N}^{\mu}.
\end{eqnarray}
In order to reduce our notation to the essentials, we  strip off all the colour indices, position arguments and Wilson lines
from ${\cal O}_{k}^{\mu}$ and represent them only by their flavour,  chiral and Dirac  structures.
The collinear quark fields moving in the direction of $\bar n$ are denoted by $\xi$,
while the anti-collinear fields for the  direction  $n$ are labelled by $\chi$.
Applying the Fierz transformation enables us to conclude  immediately  that
the emerged three evanescent operators $E_{i}^{\mu} $ vanish at $D=4$.

We now write the perturbative expansion of the QCD correlation function $\Pi^{\mu}$
in terms of the tree-level matrix elements of the collinear operators ${\cal O}_{k}^{\mu}$
\begin{eqnarray}
\Pi^{\mu} = \frac{(4 \pi)^4}{Q^6} \, \sum_{k}  \,  \sum_{\ell=0, 1}  \,
\left ( {Z_{\alpha} \alpha_s \over  4 \pi} \right )^{\ell + 2} \, A_{k}^{(\ell)} \,
\otimes \langle {\cal O}_{k}^{\mu} \rangle^{(0)} \,,
\hspace{0.8 cm}
\label{expansion of the QCD matrix element}
\end{eqnarray}
where the renormalization factor of the strong coupling is given by
$Z_{\alpha}=1 - {1 \over \epsilon} \, {\alpha_s \over  4 \pi} \, \beta_0 + {\cal O}(\alpha_s^2)$ \cite{Herzog:2017ohr,Luthe:2017ttg,Chetyrkin:2017bjc}.
The quantities $ A_{k}^{(\ell)}$ denote the bare ${\ell}$-loop on-shell QCD amplitudes
containing  $1/\epsilon$ poles of both  UV and  IR  nature.
On the other hand, the effective  matrix elements of the collinear operators
take a similar form to (\ref{expansion of the QCD matrix element})
\begin{eqnarray}
\langle {\cal O}_{k}^{\mu} \rangle = \sum_{m} \, \sum_{\ell}  \, \left ( {\alpha_s \over 4 \pi} \right )^{\ell} \,
Z_{k m}^{(\ell)} \otimes  \langle {\cal O}_{m}^{\mu} \rangle^{(0)}  \,,
\end{eqnarray}
where $Z_{k m}^{(\ell)}$ stands for the matrix kernel of UV renormalization factors
at the ${\ell}$-loop order.
Expanding further the short-distance functions $T_{k}$ in the matching relation (\ref{matching condition})
according to
$T_{k} = \sum \limits_{\ell=0}^{\infty}  \, \left ( {\alpha_s \over 4 \pi} \right )^{\ell} \, \, T_{k}^{(\ell)}$,
we are then able to deduce the  master formulae for the tree and one-loop hard  coefficients
of the two physical collinear operators
\begin{eqnarray}
T_{N(T)}^{(0)} &=&  A_{N(T)}^{(0)} \,,
\nonumber \\
T_{N(T)}^{(1)} &=&  A_{N(T)}^{(1)} + 2 \,  Z_{\alpha}^{(1)} \, A_{N(T)}^{(0)}
- \sum_{k} \, Z_{k N(T)}^{(1)}  \otimes T_{k}^{(0)},
\hspace{0.8 cm}
\end{eqnarray}
where the UV renormalization kernels  $Z_{NN}^{(1)}$ and $Z_{TT}^{(1)}$ have been worked out in  \cite{Braun:1999te}
(and verified independently by ourselves).
Following the prescription detailed in \cite{Buras:1989xd,Buras:1992tc,Dugan:1990df,Herrlich:1994kh,Beneke:2005vv},
the renormalization constants  for the evanescent operators can be determined by requiring that
the IR-finite matrix elements $\langle E_{i}^{\mu} \rangle$ vanish identically.
The peculiar evanescent-to-physical operator mixing thus generates the non-vanishing {\it finite} renormalization constants
$Z_{E_1 N}^{(1)}$ and $Z_{E_3 N}^{(1)}$, which are indispensable for correctly establishing the perturbative factorization formula
with our $D$-dimensional framework (see \cite{Beneke:2006mk,Krankl:2011gch,Anikin:2013aka,Wang:2017ijn,Gao:2021iqq,Gao:2019lta,Li:2020rcg}
for further discussions).

Employing  the definition of the  collinear matrix element for the non-local  operator ${\cal O}_{N}^{\mu}$
displayed in (\ref{definiton of the twist-3 nucleon DA}) and
\begin{eqnarray}
&& \left \langle 0 \left |   \epsilon_{i j k} \, \left [  u_{i}^{\uparrow}(\tau_1 \, n) \,  C \gamma_{\perp \alpha} \slashed{n} \,
u_{j}^{\downarrow}(\tau_2 \, n) \right ]  \gamma_{\perp}^{\alpha}
\slashed{n}  \, d_{k}^{\uparrow}(\tau_3 \, n) \right | N^{\uparrow} (P) \right  \rangle
\nonumber \\
&& = 2 \, f_N(\mu_F) \, (n \cdot P)  \, \slashed{n} \, N^{\uparrow}(P)
\int [{\cal D} x] \, {\rm exp} \left [ -  i  \, n \cdot P   \sum_{i=1}^{3} x_i \, \tau_i \right ]
\nonumber \\
&& \hspace{0.5 cm}  \varphi_T(x_i, \mu_F)  \,,
\end{eqnarray}
where the three finite-length  Wilson lines are not explicitly  written out,
we can readily express the desired short-distance coefficient function  ${\mathbb T}_{N}$ entering the factorized expression
(\ref{hard-collinear factorization formula}) as follows
\begin{eqnarray}
&& {\mathbb T}_{N}(x_1, x_2, x_3,  y_1, y_2, y_3,  Q^2, \mu_F)
\nonumber \\
&& = T_{N}(x_1, x_2, x_3,  y_1, y_2, y_3,  Q^2, \mu_F) + \left ({1 \over 4} \right )  \,
\nonumber \\
&&   \hspace{0.3 cm} \times \, \bigg [ T_{T}(x_1, x_3, x_2,  y_1, y_3, y_2,  Q^2, \mu_F)
+  (x_1 \leftrightarrow x_3)
\nonumber \\
&&  \hspace{0.8 cm} + \,  (y_1 \leftrightarrow y_3) + (x_1 \leftrightarrow x_3, \, y_1 \leftrightarrow y_3)   \bigg ].
\end{eqnarray}
An interesting relation for the two nucleon distribution amplitudes of twist-three
$\varphi_T(x_1, x_2, x_3, \mu_F)  =  \left ({1 / 2} \right ) \,
\left [ \varphi_N(x_1, x_3, x_2, \mu_F)  + \varphi_N(x_2, x_3, x_1, \mu_F)   \right ]$
due to isospin symmetry \cite{Chernyak:1984bm,King:1986wi,Chernyak:1987nv}
has been taken into account.
Substituting the newly determined one-loop hard function into the perturbative factorization formula
(\ref{hard-collinear factorization formula}) and evaluating  the convolution integral
with the asymptotic nucleon distribution amplitude $\varphi_N$ gives rise to
\begin{widetext}
\begin{eqnarray}
F_1^{\rm Asy}(Q^2) &=&  \frac{(4  \pi  \alpha_s)^2} {Q^4} \, 75 \, f_N^2  \,
\left ( {N_c +1 \over 2 \, N_c} \right )^2 \,
\bigg [ Q_u \, \bigg \{  1 +  { \alpha_s \over4 \pi } \,
\bigg [ \left ( 2 \, \beta_0 + C_F \right ) \, \ln{\mu_F^2 \over Q^2}  + {46 \over 3} \, \beta_0
 -  C_F  \, \left ( {7171 \over 20} + 544 \, \zeta_3 - 864 \, \zeta_5 \right )
\nonumber \\
&& \hspace{4.5 cm} + \, 2 \, C_A \, \left (  {107 \over 5}  + {2174 \over 15} \, \zeta_3 - {424 \over 3} \, \zeta_5 \right )
+ {8 \, N_c^2 \over N_c +1} \, \left (  {76 \over 15} + {293 \over 15} \, \zeta_3  - {112 \over 3} \, \zeta_5  \right )  \bigg ]  \bigg \}
\nonumber \\
&&  + \, 2 \, Q_d  \, \bigg \{  1 +  { \alpha_s \over4 \pi }  \,
\bigg  [  \left ( 2 \, \beta_0 + C_F \right ) \, \ln{\mu_F^2 \over Q^2}
+ { 65 \over 6} \, \beta_0  - C_F \, \left (  {2053 \over 40} + {14 \over 5} \, \zeta_3  - 24 \,  \zeta_5   \right )
 + C_A \, \left (  {73 \over 5} + {226 \over 15} \, \zeta_3  - {56 \over 3} \, \zeta_5 \right )
 \nonumber \\
&& \hspace{2.5 cm}  -  \, {2 \over 3} \, {N_c^2 \over N_c +1} \, \left ( 13 + 23 \, \zeta_3  - 20 \, \zeta_5 \right ) \bigg ]
\bigg \}   \bigg ] \,,
\end{eqnarray}
\end{widetext}
where the Mathematica package {\tt PolyLogTools} \cite{Duhr:2019tlz} turns out to be highly beneficial
for  handling  the encountered four-fold integrals {\it analytically}.
Including the subleading terms in the conformal expansion of the twist-three distribution amplitude  $\varphi_N$
results in the length expression for the non-asymptotic correction,
which are  presented explicitly in the Supplemental Material for completeness.
It is straightforward to  demonstrate that our NLO QCD computation of the Dirac nucleon form factor
based upon  the hard-collinear factorization formula (\ref{hard-collinear factorization formula})
is indeed independent of the factorization scale $\mu_F$ at the ${\cal O}(\alpha_s^3)$ accuracy,
by employing the one-loop evolution equation of the leading-twist nucleon distribution amplitude $\varphi_N$ \cite{Braun:1999te}.
Achieving the next-to-leading-logarithmic (NLL) resummation of $\ln (Q^2 / \Lambda_{\rm QCD}^2 )$
in the factorized expression (\ref{hard-collinear factorization formula}) necessitates
the complete two-loop RG evolution  equation of the nucleon distribution amplitude $\varphi_N$,
which is,  unfortunately,  not yet determined thus far
(except for the two-loop anomalous dimension of the normalization constant $f_N$
in the so-called KM scheme \cite{Krankl:2011gch,Gracey:2012gx}).

%
\section{Numerical Analysis}
%

We are now in a position to explore the phenomenological implication of the newly determined NLO QCD correction
to the Dirac nucleon form factor, including also an elaborate comparison with the available experimental measurements.
To achieve this goal, we  first specify the fundamental non-perturbative input,
namely, the twist-three nucleon distribution amplitude,
appearing in the hard-collinear factorization formula of the Dirac nucleon form factor,
which has been extensively investigated with the method of QCD sum rules (QCDSR)
\cite{Chernyak:1984bm,Chernyak:1987nt,Chernyak:1987nv,Chernyak:1987nu,King:1986wi,Gari:1986ue,Gari:1986dr,Stefanis:1992nw,Bergmann:1993rz},
with the LCSR technique \cite{Braun:2005be,Braun:2006hz,Lenz:2009ar,Anikin:2013aka},
and with the lattice QCD simulation \cite{QCDSF:2008qtn,Braun:2014wpa,Bali:2015ykx,RQCD:2019hps}.
Three sample models of the initial condition $\varphi_N(x_i, \mu_0)$
at a reference scale  $\mu_{0}^2 = 2.0 \, {\rm GeV^2}$,
labelled as {\tt ABO1} \cite{Anikin:2013aka}, {\tt LAT19} \cite{RQCD:2019hps}, and {\tt COZ} \cite{Chernyak:1987nt},
will be employed in the subsequent numerical exploration for illustration purposes.
The construction of  {\tt ABO1} is achieved  by comparing the NLO LCSR predictions of
nucleon electromagnetic form factors \cite{Anikin:2013aka} with the experimental data points
\cite{Arrington:2007ux,CLAS:2008idi,JeffersonLabHallA:2001qqe,Punjabi:2005wq,Puckett:2010ac,JeffersonLaboratoryE93-038:2005ryd,Riordan:2010id}.
The normalization  constant and shape parameters in the {\tt LAT19} model
are obtained from the improved $N_f = 2 +1$ lattice QCD analysis with dynamical clover fermions \cite{Bruno:2014jqa}.
The parameter set {\tt COZ} is determined by the tree-level QCDSR computation with the inclusion of
the power-suppressed contributions up to the dimension-six vacuum condensates \cite{Chernyak:1987nt}.
Furthermore, we will tacitly take the renormalization scale $\nu$ of the strong coupling $\alpha_s$
and the factorization scale  $\mu_F$ characterizing  the short-distance fluctuations of
off-shell quark and gluon  fields    as $\nu^2 =\mu_F^2 = \langle z \rangle \, Q^2$
with $ 1/6 \leq \langle z \rangle  \leq 1/2$,
which corresponds to the expectation value of the maximum quark/gluon virtuality in the LO Feynman graphs
(see \cite{Stefanis:1987vr,Li:1992ce,Bolz:1994hb,Stefanis:1997zyh,He:2006ud,He:2006vz,Lu:2009cm,Li:2010nn,Li:2012nk,Li:2013xna,Li:2014xda}
for further discussions on choices of the factorization scale).

\begin{figure}[htp]
\includegraphics[width=0.45 \textwidth]{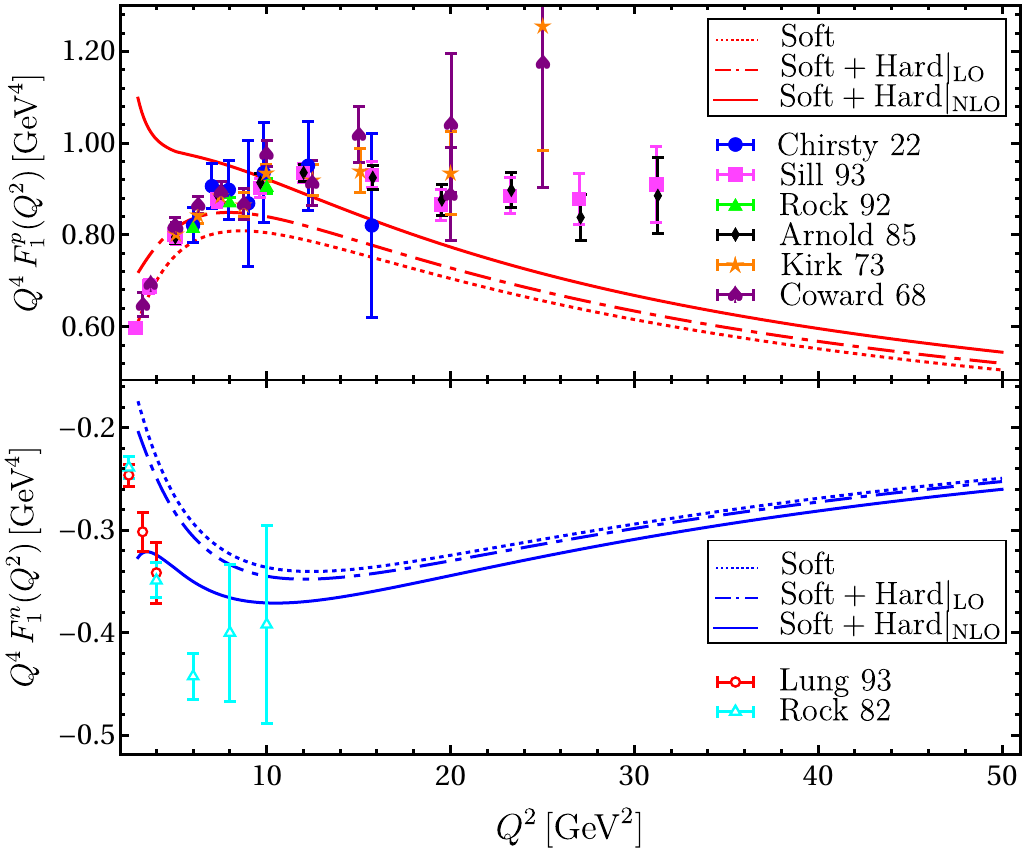}
\caption{Theory predictions for the  Dirac form factors of both the proton and the neutron
from the soft Feynman mechanisms (dotted curves),
and from the combinations of the above-mentioned  contributions with
the so-called hard-scattering mechanisms  at LO (dashed-dotted curves) and at NLO (solid curves).
We further display the  experimental data points  of the Dirac proton form factor
(Christy 22 \cite{Christy:2021snt}, Sill 93 \cite{Sill:1992qw}, Rock 92 \cite{Rock:1991jy},
Arnold 85 \cite{Arnold:1986nq},  Kirk 73 \cite{Kirk:1972xm}, Coward 68 \cite{Coward:1967au})
and of the Dirac neutron form factor (Lung 93 \cite{Lung:1992bu}, Rock 82 \cite{Rock:1982gf}) for an exploratory comparison.}
\label{fig:predictions-for-the-seperate-contributions}
\end{figure}

In order to develop a transparent understanding of  distinct dynamical mechanisms dictating
the Dirac electromagnetic nucleon form factors at experimentally accessible momentum transfers,
we display explicitly in Figure \ref{fig:predictions-for-the-seperate-contributions}
 the yielding predictions for the  next-to-leading power soft contributions
from the LCSR approach at ${\cal O}(\alpha_s)$ \cite{Anikin:2013aka}
and for the (competing) hard-gluon-exchange contributions from the perturbative factorization formalism
at LO  and at NLO accomplished in this Letter, by taking the {\tt ABO1} model as our default choice.
The available experimental measurements of these two transition form factors
\cite{Christy:2021snt,Sill:1992qw,Rock:1991jy,Arnold:1986nq,Kirk:1972xm,Coward:1967au, Lung:1992bu,Rock:1982gf}
are further displayed for an exploratory comparison.
It is perhaps worth mentioning that there is no  double counting
when adding the NLO LCSR computations of the soft contributions
on top of the  hard perturbative QCD contributions,
because of the fact that the tree-level hard scattering mechanism
can be only generated at the level of ${\cal O}(\alpha_s^2)$
while the soft (end-point) contributions are evaluated with the LCSR method at the one-loop accuracy.
It is evident from Figure \ref{fig:predictions-for-the-seperate-contributions} that
the (formally) leading-power hard-scattering contributions can shift the NLO LCSR predictions
for both the Dirac proton and neutron form factors  by an amount of  approximately $(5-20) \, \%$
in the kinematic region $Q^2 \in [5.0, \, 50.0] \, {\rm GeV^2}$.
In particular, the newly computed  radiative corrections to the hard-scattering kernels
in the QCD factorization formula (\ref{hard-collinear factorization formula})
can bring about   enormous impacts on the corresponding tree-level predictions
for a wide range of the kinematic domain.
We further verify that this  enlightening  pattern of the perturbative $\alpha_s$ expansion  remains  unchanged for
both the {\tt LAT19}  and  the {\tt COZ} models, thus justifying  the extraordinary phenomenological significance
of carrying out the complete one-loop computations for the nucleon electromagnetic form factors.

\begin{figure}[htp]
\includegraphics[width=0.45 \textwidth]{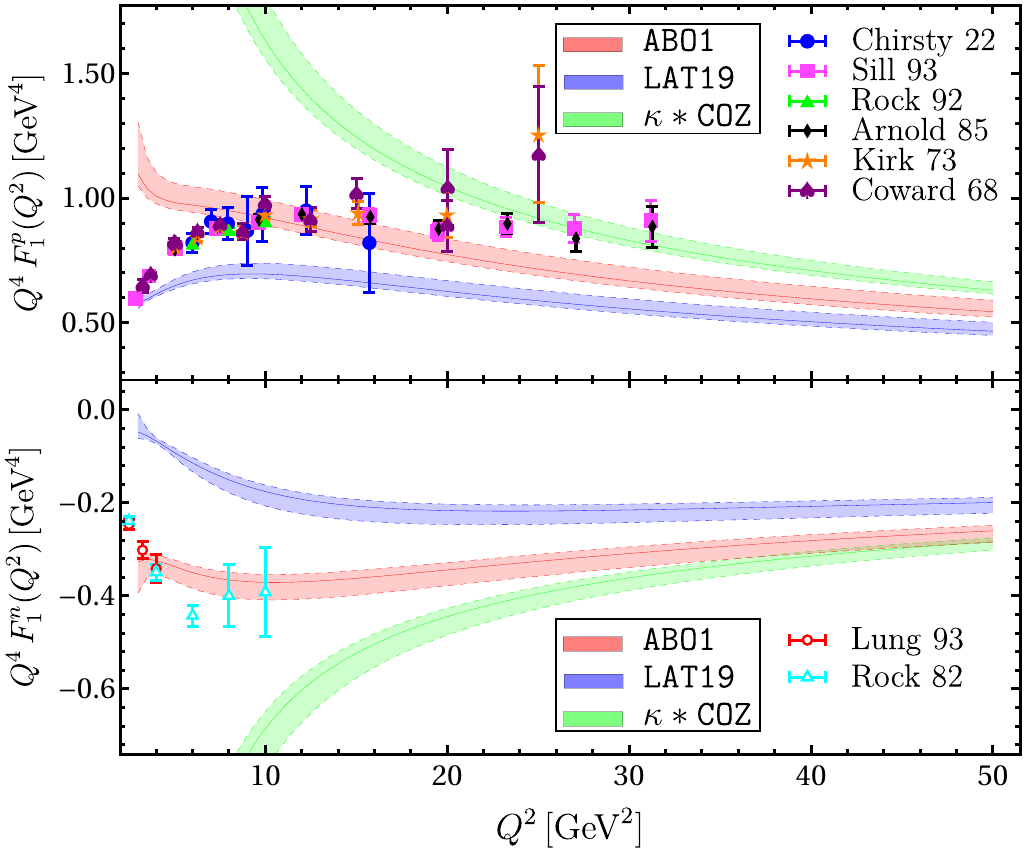}
\caption{Theory predictions for  the  Dirac form factors of both the proton and the neutron
from the three sample models ({\tt ABO1} \cite{Anikin:2013aka}, {\tt LAT19} \cite{RQCD:2019hps},
and {\tt COZ} \cite{Chernyak:1987nt}) of the nucleon distribution amplitudes,
obtained by combining the  soft (end-point) contributions from \cite{Anikin:2013aka}
with the NLO QCD predictions of the hard-scattering contributions.
The perturbative uncertainties from varying the factorization scale $\mu_F$ are indicated by the colour bands.
Note that the achieved  results from the {\tt COZ} model are rescaled with a factor of $\kappa=0.5$.
The references to  the quoted experimental  data points are already summarized
in the caption of Figure \ref{fig:predictions-for-the-seperate-contributions}.}
\label{fig:model-predictions-vs-exp-data}
\end{figure}

We proceed to present in Figure \ref{fig:model-predictions-vs-exp-data} our  theory predictions
for the Dirac nucleon form factors with  the three sample models
{\tt ABO1} \cite{Anikin:2013aka}, {\tt LAT19} \cite{RQCD:2019hps}, and {\tt COZ} \cite{Chernyak:1987nt}
for nucleon distribution amplitudes,
which confront  with  the experimental data points spanning more than half a century
\cite{Christy:2021snt,Sill:1992qw,Rock:1991jy,Arnold:1986nq,Kirk:1972xm,Coward:1967au, Lung:1992bu,Rock:1982gf}.
Inspecting the numerical features of these state-of-the-art results  implies that
the yielding theory predictions  with   the LCSR-inspired  {\tt ABO1} model can lead to
a reasonably good description of  the available experimental data
for both the proton and neutron's Dirac form factors at intermediate momentum transfers.
On the contrary, the predicted  electromagnetic form factors $F_1^{p}(Q^2)$ and $F_1^{n}(Q^2)$
with the QCDSR-based {\tt COZ} model  for   the nucleon distribution amplitudes
turn out to be significantly above their experimental values  in the entire kinematic region.
This unexpected observation can be attributed to the fact that the QCDSR method
based upon the local operator product expansion
tends to overestimate the hadronic matrix elements of higher conformal operators considerably  \cite{Bakulev:1991ps,Mikhailov:1991pt,Braun:2001tj,Braun:2005be,Braun:2006hz,Lenz:2009ar}.
The distinctive snapshot of the well-separated error bands
for the two   parameter sets {\tt ABO1} and {\tt LAT19}
can be  mainly traced backed to the very absence of the lattice results for the shape parameters
$\eta_{10}$ and $\eta_{11}$  appearing in  the twist-four nucleon distribution amplitudes $\Phi_4$ and $\Psi_4$,
whose effects dominate the soft non-factorizable contributions to the nucleon  form factors
at intermediate momentum transfers in the LCSR framework \cite{Braun:2001tj,Braun:2006hz,Anikin:2013aka}.
Alternatively,  the observed discrepancies  between the pink and blue uncertainty bands
arise --- to a large extent --- from  non-asymptotic corrections to the twist-four pieces of three-quark configurations
with different helicity structure in the LCSR analysis,
thus elucidating the importance of quark orbital angular momentum components inside the composite nucleon state.
In short, there are actually no genuine tensions for the obtained theory  predictions
with the parameter sets   {\tt ABO1} and {\tt LAT19}, upon adopting the same numerical results
for the two twist-four shape parameters $\eta_{10}$ and $\eta_{11}$ discussed above.

\begin{figure}[htp]
\includegraphics[width=0.45 \textwidth]{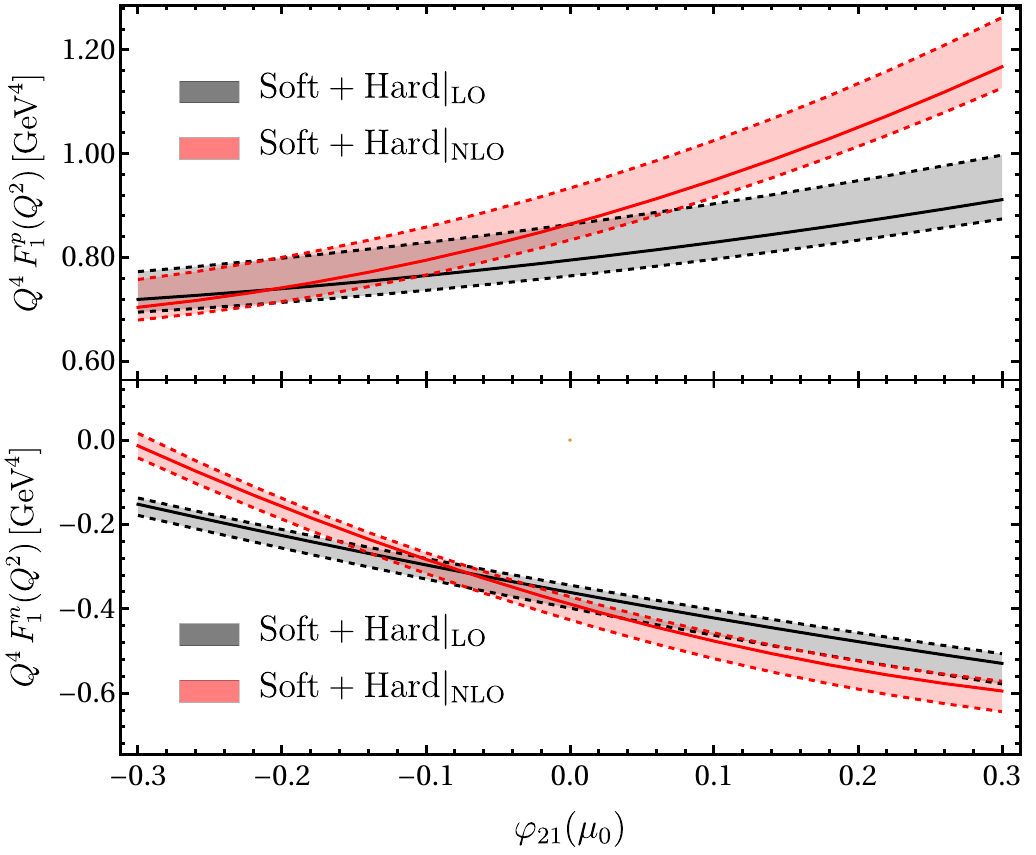}
\caption{Theory predictions for the $\varphi_{21}(\mu_0)$ dependence of the Dirac nucleon form factors
at $Q^2 = 15.0 \, {\rm GeV^2}$ obtained by combining the power-suppressed soft contributions (from \cite{Anikin:2013aka})
with the counterpart hard-scattering contributions at LO (grey bands) and at NLO (pink bands).
The remaining parameters for the nucleon distribution amplitudes are identical to the ones employed in the {\tt ABO1} model.}
\label{fig:shape-parameter-dependence}
\end{figure}

We now turn to address  the phenomenological impact of extracting the essential shape parameters
for  the twist-three nucleon distribution amplitude $\varphi_N$ with the NLO QCD computations of
the hard-gluon-exchange contributions to the Dirac nucleon form factors.
For the purpose of illustration, we display in Figure \ref{fig:shape-parameter-dependence}
the intrinsic dependence  of the $\gamma^{\ast} N \to N$ transition form factors
on the currently poorly constrained parameter $\varphi_{21}(\mu_0)$,
while adopting the other non-perturbative parameters from the  {\tt ABO1} model.
We can readily observe  that the inclusion of the  higher-order QCD correction to the short-distance coefficient function
in the factorization formula (\ref{hard-collinear factorization formula})
can indeed profoundly  enhance the sensitivity of the  nucleon electromagnetic form factors
with regard to  the local moment $\varphi_{21}(\mu_0)$ of the nucleon distribution amplitude.
We finally mention in passing that here  we restrict ourselves to  precision QCD predictions of the space-like nucleon
electromagnetic form factors and reserve their theoretically  challenging time-like counterparts \cite{Denig:2012by}
for our subsequent work.

\section{Conclusions}

In conclusion, we have carried out, for the first time,  the complete NLO QCD computations
of the proton and neutron's Dirac form factors
by taking advantage of the hard-collinear factorization theorem  rigorously.
Adopting  three sample models for the nucleon distribution amplitudes,
we demonstrated explicitly that the newly determined NLO corrections
to the hard-scattering contributions can yield  the substantial phenomenological impacts
on the counterpart LO predictions in the kinematic region $5.0 \, {\rm GeV}^2 \leq  Q^2  \leq 50.0 \, {\rm GeV}^2$.
In particular, taking into account the higher-order QCD correction
turned out to be highly beneficial for an improved extraction of the shape parameters
of the nucleon distribution amplitudes.
Extending our factorization formalism to the hadron form factors
describing the electroproductions of the $\Delta$-resonance  and  the negative-parity $N^{\ast}(1535)$
and  to the exclusive semileptonic $\Lambda_b \to p \ell \bar \nu_{\ell}$  decay form factors
can be then naturally expected for the sake of exploring the many facets of  QCD dynamics
dictating the baryon transition matrix elements.

%
\begin{acknowledgments}
\section*{Acknowledgements}

We are grateful to   Vladimir M. Braun for illuminating discussions and
for  valuable  comments/suggestions  on the manuscript.
This research is supported in part by the  National Natural Science Foundation of China
with Grant No.  12075125.

\end{acknowledgments}


\section{Note Added}

While this paper was being finalized  the work \cite{Chen:2024fhj} appeared,
in which  the light-cone projection on the twist-three collinear operators
was implemented {\it before} renormalizing the $7$-point QCD amplitude $\Pi_{\mu}$
without including  an important class of contributions from   evanescent operators.
By ignoring  finite renormalization terms due to the evanescent-to-physical operator mixing,
the premature strategy to extract the short-distance matching coefficients described in  \cite{Chen:2024fhj}
cannot be justified already at the one-loop accuracy,  as widely discussed in the factorization analyses
of  light-flavour hadron transition form factors \cite{Anikin:2013aka,Krankl:2011gch,Wang:2017ijn,Gao:2021iqq}
and of heavy-quark hadron decay matrix elements \cite{Buras:1989xd,Buras:1992tc,Dugan:1990df,Herrlich:1994kh,Beneke:2005vv}.
This observation can be further verified from the very fact that
their (approximate) numerical values of the coefficients $c_2$ and $d_2$ collected in Table I of \cite{Chen:2024fhj}
are in evident contradiction with the (exact) analytical results displayed in the Supplementary Material of this Letter.

Moreover, the authors of \cite{Chen:2024fhj} {\it did  not}
take into account the numerically dominant soft non-factorizable contributions
to the Dirac nucleon form factors at  experimentally accessible momentum transfers $Q^2$
in their phenomenological investigations.
Unsurprisingly, they were then led to  an incorrect conclusion that
the resulting theory predictions for the Dirac form factors of both the proton and the neutron
with the non-perturbative parameters of the twist-three nucleon distribution amplitude
from the recent lattice QCD simulations \cite{Braun:2014wpa,RQCD:2019hps}
cannot accommodate the available experimental data.
An elaborate discussion on this longstanding issue  has been  documented
in a large number of excellent review articles
(see for instance \cite{Radyushkin:1984wq,Bolz:1994hb,Stefanis:1997zyh,Perdrisat:2006hj,Aznauryan:2012ba,Punjabi:2015bba,Pire:2021hbl}).

%
\appendix

\begin{widetext}

\section{SUPPLEMENTAL MATERIAL}

Inserting the conformal expansion of the twist-three nucleon distribution amplitude
into the one-loop factorization formula of the Dirac nucleon form factor
and performing the four-fold  integrals over  the momentum fractions {\it analytically} yields
(for the physical SU(3) colour group)
\begin{eqnarray}
F_1(Q^2) =  \frac{(4  \pi  \alpha_s)^2} {Q^4} \, {100 \over 3} \,
\left [ f_N(\mu_F) \right ]^2  \,  \sum_{m, \, k, \, m^{\prime}, \,  k^{\prime}} \,
 \varphi_{m k}(\mu_F)  \, \varphi_{m^{\prime} k^{\prime}}(\mu_F)  \,
\left [ Q_u \,\,    {\cal U}_{m k}^{m^{\prime} k^{\prime}}(\mu_F)
+ Q_d \,\,   {\cal D}_{m k}^{m^{\prime} k^{\prime}}(\mu_F)  \right ]\,,
\end{eqnarray}
where the  yielding  coefficient matrices  ${\cal U}$ and ${\cal D}$  can be further  expanded perturbatively in QCD
\begin{eqnarray}
{\cal U}_{m k}^{m^{\prime} k^{\prime}} =  {\cal N}_{m k}^{m^{\prime} k^{\prime}} \,
\left \{  1 +  \left ( {\alpha_s  \over 4 \, \pi}  \right )  \,
\left [  \left ( 2 \, \beta_0 + \left (\gamma_{m k}^{(1)} + \gamma_{m^{\prime} k^{\prime}}^{(1)}  \right ) \,  C_F  \right ) \,
\ln{\mu_F^2 \over Q^2}   + {\cal R}_{m k}^{m^{\prime} k^{\prime}} \right ]
+ {\cal O}(\alpha_s^2) \right \}  \,,
\nonumber \\
{\cal D}_{m k}^{m^{\prime} k^{\prime}} =  {\cal \widetilde{N}}_{m k}^{m^{\prime} k^{\prime}} \,
\left \{  1 +  \left ( {\alpha_s  \over 4 \, \pi}  \right )  \,
\left [  \left ( 2 \, \beta_0 + \left (\gamma_{m k}^{(1)} + \gamma_{m^{\prime} k^{\prime}}^{(1)}  \right ) \,  C_F  \right ) \,
\ln{\mu_F^2 \over Q^2}   + {\cal \widetilde{R}}_{m k}^{m^{\prime} k^{\prime}} \right ]
+ {\cal O}(\alpha_s^2) \right \}  \,.
\end{eqnarray}
It is straightforward to verify that the flavour-independent coefficients $\gamma_{m^{(\prime)} k^{(\prime)}}^{(1)}$
coincide with the one-loop anomalous dimensions (up to an overall minus sign)
of the local moments $f_N \, \varphi_{m^{(\prime)} k^{(\prime)}}$
for the twist-three nucleon distribution amplitude $\varphi_N$,
whose explicit expressions for   $m^{(\prime)} \leq 2$ used in our computation are given by \cite{Peskin:1979mn,Tesima:1981ud,Stefanis:1994zd,Bergmann:1999ud,Braun:1999te}
\begin{eqnarray}
\gamma_{0 0}^{(1)} = {1 \over 2} \,, \qquad \gamma_{1 0}^{(1)} = {13 \over 6}  \,,  \qquad \gamma_{1 1}^{(1)} = {5 \over 2}  \,,
\qquad
\gamma_{2 0}^{(1)} = {19 \over 6}\,,   \qquad \gamma_{2 1}^{(1)} = {23 \over 6}\,,  \qquad   \gamma_{2 2}^{(1)} = 4 \,.
\end{eqnarray}
The normalization (tree-level) coefficients ${\cal N}$ and ${\cal \widetilde{N}}$
together with the NLO kernels ${\cal R}$ and ${\cal \widetilde{R}}$ can be written as
\begin{align}
\left\{ \mathcal{N}_{00}^{00}, \, \mathcal{N}_{00}^{10}, \,  \mathcal{N}_{00}^{11}, \,  \mathcal{N}_{00}^{20},  \,
\mathcal{N}_{00}^{21}, \,  \mathcal{N}_{00}^{22} \right \}
&= \left \{1, \frac{49}{6},  \,  \frac{7}{2},  \,  \frac{427}{20},   \,  \frac{35}{4},  \,  \frac{16}{5} \right\},
\nonumber \\
\left\{ \mathcal{N}_{10}^{10}, \,  \mathcal{N}_{10}^{11}, \,  \mathcal{N}_{10}^{20}, \,  \mathcal{N}_{10}^{21}, \,
\mathcal{N}_{10}^{22}, \mathcal{N}_{11}^{11}, \,  \mathcal{N}_{11}^{20},  \,  \mathcal{N}_{11}^{21}, \,  \mathcal{N}_{11}^{22} \right\}
&=  \left \{ \frac{1519}{9}, \,  \frac{343}{9}, \,  \frac{539}{6}, \,  \frac{637}{4}, \,  \frac{49}{4}, \,  \frac{539}{9}, \,
\frac{931}{10}, \,  \frac{539}{12}, \,  \frac{427}{60} \right \},
\nonumber \\
\left \{ \mathcal{N}_{20}^{20}, \, \mathcal{N}_{20}^{21}, \, \mathcal{N}_{20}^{22}, \, \mathcal{N}_{21}^{21}, \, \mathcal{N}_{21}^{22}, \, \mathcal{N}_{22}^{22} \right \}
&=  \left \{ \frac{23471}{100}, \, \frac{3381}{40}, \,\frac{6069}{200}, \, \frac{637}{4}, \, \frac{287}{40}, \, \frac{779}{100} \right\},
\nonumber \\
\left \{ \widetilde{\mathcal{N}}_{00}^{00}, \, \widetilde{\mathcal{N}}_{00}^{10}, \, \widetilde{\mathcal{N}}_{00}^{11}, \, \widetilde{\mathcal{N}}_{00}^{20}, \, \widetilde{\mathcal{N}}_{00}^{21}, \, \widetilde{\mathcal{N}}_{00}^{22}\right\}
&=  \left \{ 2, \,  -\frac{49}{6}, \,  -\frac{7}{2}, \,  \frac{161}{20}, \,  -\frac{35}{4}, \,  \frac{1}{10} \right \},
\nonumber \\
\left \{ \widetilde{\mathcal{N}}_{10}^{10}, \, \widetilde{\mathcal{N}}_{10}^{11}, \, \widetilde{\mathcal{N}}_{10}^{20}, \, \widetilde{\mathcal{N}}_{10}^{21}, \, \widetilde{\mathcal{N}}_{10}^{22}, \, \widetilde{\mathcal{N}}_{11}^{11}, \, \widetilde{\mathcal{N}}_{11}^{20}, \, \widetilde{\mathcal{N}}_{11}^{21}, \, \widetilde{\mathcal{N}}_{11}^{22}\right\}
&=  \left\{\frac{1568}{9}, \, -\frac{343}{9}, \, -\frac{539}{6}, \, \frac{637}{4}, \, -\frac{49}{4}, \, \frac{196}{9}, \, \frac{49}{5}, \, -\frac{539}{12}, \, \frac{161}{60} \right \}, 
\nonumber \\
\left \{ \widetilde{\mathcal{N}}_{20}^{20}, \, \widetilde{\mathcal{N}}_{20}^{21}, \, \widetilde{\mathcal{N}}_{20}^{22}, \, \widetilde{\mathcal{N}}_{21}^{21}, \, \widetilde{\mathcal{N}}_{21}^{22}, \, \widetilde{\mathcal{N}}_{22}^{22} \right \}
&=   \left \{ \frac{1372}{25}, \, -\frac{3381}{40}, \, \frac{987}{200}, \, \frac{637}{4}, \, -\frac{287}{40}, \, \frac{43}{25} \right \},
\end{align}
and
\begin{align}
\mathcal{R}_{00}^{00} & = -\frac{1807}{15} + \frac{7438}{15}  \, \zeta _3 - 368 \, \zeta _5,
& \widetilde{\mathcal{R}}_{00}^{00} = & \frac{1601}{30} + \frac{209}{30}  \,\zeta _3 + 6 \,\zeta _5,
\nonumber \\
\mathcal{R}_{00}^{10} & = \frac{1025389 }{7560} + \frac{3203}{10} \,\zeta _3 -400 \,\zeta _5,
& \widetilde{\mathcal{R}}_{00}^{10} = & \frac{1025389}{7560} + \frac{3203 }{10}\, \zeta _3 -400 \, \zeta _5,
\nonumber \\
\mathcal{R}_{00}^{11} & = -\frac{12211 }{216} + \frac{5161}{18} \,\zeta _3 - \frac{520 }{3} \, \zeta _5,
& \widetilde{\mathcal{R}}_{00}^{11} = & \frac{106891 }{1080} - \frac{11953 }{90} \,\zeta _3 + \frac{368 }{3} \, \zeta _5,
\nonumber  \\
\mathcal{R}_{00}^{20} & =\frac{109181473 }{461160} + \frac{18937349}{38430} \, \zeta _3 - \frac{298264 }{427} \, \zeta _5,
& \widetilde{\mathcal{R}}_{00}^{20} = & \frac{12924161 }{173880} - \frac{244199 }{14490} \,\zeta _3 + \frac{7120 }{161} \,\zeta _5,
\nonumber  \\
\mathcal{R}_{00}^{21} & = -\frac{47659 }{12600} - \frac{179843}{1050} \, \zeta _3 + \frac{10928 }{35} \, \zeta _5,
& \widetilde{\mathcal{R}}_{00}^{21} = & \frac{15061 }{12600} - \frac{179843 }{1050} \,\zeta _3 + \frac{10928 }{35} \, \zeta _5,
\nonumber \\
\mathcal{R}_{00}^{22} & = \frac{7769413 }{34560} + \frac{1924849 }{2880} \, \zeta _3 - \frac{3543}{4}  \, \zeta _5,
& \widetilde{\mathcal{R}}_{00}^{22} = & -\frac{1421329 }{1080} - \frac{144079 }{90} \, \zeta _3 + 3360 \,\zeta _5,
 \\
\nonumber \\
\mathcal{R}_{10}^{10} & = \frac{11754337 }{33480} + \frac{151409}{217} \, \zeta _3 - \frac{223704 }{217} \, \zeta _5,
& \widetilde{\mathcal{R}}_{10}^{10} =& \frac{5837081 }{30240} + \frac{160177 }{560} \, \zeta _3 - \frac{22419 }{56} \, \zeta _5,
\nonumber \\
\mathcal{R}_{10}^{11} & = \frac{23519059 }{52920} + \frac{447161}{490} \, \zeta _3 - \frac{9600 }{7} \, \zeta _5,
& \widetilde{\mathcal{R}}_{10}^{11} = & \frac{24043219 }{52920} + \frac{447161 }{490} \, \zeta _3 - \frac{9600 }{7} \,\zeta _5,
\nonumber \\
\mathcal{R}_{10}^{20} & = \frac{23177779 }{55440} + \frac{1590773}{1925} \, \zeta _3 - \frac{476496 }{385} \,\zeta _5,
& \widetilde{\mathcal{R}}_{10}^{20} = & \frac{23332339 }{55440} + \frac{1590773 }{1925} \, \zeta _3 - \frac{476496 }{385} \, \zeta _5,
\nonumber \\
\mathcal{R}_{10}^{21} & = -\frac{48986191 }{196560} - \frac{447763}{455} \, \zeta _3 + \frac{19632 }{13} \, \zeta _5,
& \widetilde{\mathcal{R}}_{10}^{21} = & \frac{6568573 }{28080} + \frac{129473 }{455} \, \zeta _3 - \frac{38184 }{91} \, \zeta _5,
\nonumber \\
\mathcal{R}_{10}^{22} & = -\frac{2665223 }{3024} - \frac{439976}{175} \, \zeta _3 + \frac{136272 }{35} \, \zeta _5,
& \widetilde{\mathcal{R}}_{10}^{22} = & -\frac{13315363 }{15120} - \frac{439976 }{175} \, \zeta _3 + \frac{136272 }{35} \, \zeta _5,
\nonumber \\
\mathcal{R}_{11}^{11} & = -\frac{457259 }{7560} - \frac{38366}{385} \, \zeta _3 + \frac{3048 }{11} \, \zeta _5,
& \widetilde{\mathcal{R}}_{11}^{11} = & \frac{770971 }{7560} + \frac{17807 }{140} \,\zeta _3 - 141 \, \zeta _5,
\nonumber \\
\mathcal{R}_{11}^{20} & = \frac{618805333 }{861840} + \frac{19751197}{11970} \, \zeta _3 - \frac{47168 }{19} \, \zeta _5,
& \widetilde{\mathcal{R}}_{11}^{20} = & \frac{56716139 }{90720} + \frac{367657 }{252} \, \zeta _3 - \frac{14942 }{7} \, \zeta _5,
\nonumber \\
\mathcal{R}_{11}^{21} & = -\frac{88513517 }{166320} - \frac{614882}{385} \,\zeta _3 + \frac{191760 }{77} \, \zeta _5,
& \widetilde{\mathcal{R}}_{11}^{21} = & -\frac{86900717 }{166320} - \frac{614882 }{385} \,\zeta _3 + \frac{191760 }{77} \,\zeta _5,
\nonumber \\
\mathcal{R}_{11}^{22} & = \frac{140188409 }{43920} + \frac{7011784}{915} \, \zeta _3 - \frac{720432}{61}  \, \zeta _5,
& \widetilde{\mathcal{R}}_{11}^{22} = & \frac{9773647 }{16560} + \frac{61132 }{69} \, \zeta _3 - \frac{33408 }{23} \, \zeta _5,
 \\
\nonumber \\
\mathcal{R}_{20}^{20} & = \frac{700662719 }{603540} + \frac{46755193}{16765} \, \zeta _3 - \frac{14151960 }{3353} \, \zeta _5,
& \widetilde{\mathcal{R}}_{20}^{20} = & -\frac{16413337}{40320} - \frac{2824781}{1960} \,\zeta _3 + \frac{430209}{196} \, \zeta _5,
\nonumber \\
\mathcal{R}_{20}^{21} & = -\frac{88263179 }{1043280} - \frac{1422557 }{2415} \, \zeta _3 + \frac{145392 }{161} \, \zeta _5,
& \widetilde{\mathcal{R}}_{20}^{21} = & -\frac{82557899}{1043280} - \frac{1422557}{2415} \, \zeta _3 + \frac{145392 }{161} \,\zeta _5,
\nonumber \\
\mathcal{R}_{20}^{22} & = -\frac{25422485051}{13109040} - \frac{56101522}{10115} \, \zeta _3 + \frac{17071440}{2023} \,\zeta _5,
& \widetilde{\mathcal{R}}_{20}^{22} = & -\frac{6503553877 }{2131920} - \frac{14010002}{1645} \,\zeta _3 + \frac{4264896}{329} \,\zeta _5,
\nonumber \\
\mathcal{R}_{21}^{21} & = \frac{673585231}{589680} + \frac{249294}{91} \,\zeta _3 - \frac{376056}{91} \, \zeta _5,
& \widetilde{\mathcal{R}}_{21}^{21} = & -\frac{31772033 }{65520} - \frac{106337}{65} \,\zeta _3 + \frac{229140}{91} \, \zeta _5,
\nonumber \\
\mathcal{R}_{21}^{22} & = \frac{133976299}{15120} + \frac{33436042}{1435}  \,\zeta _3 - \frac{10160496}{287}  \, \zeta _5,
& \widetilde{\mathcal{R}}_{21}^{22} = & \frac{5497544099}{619920} + \frac{33436042}{1435} \, \zeta _3
-\frac{10160496}{287}  \, \zeta _5,
\nonumber \\
\mathcal{R}_{22}^{22} & = -\frac{11849422619}{2523960} - \frac{48634536}{3895}  \,\zeta _3 + \frac{14908320}{779}  \, \zeta _5,
& \widetilde{\mathcal{R}}_{22}^{22} = & \frac{280771277}{557280} + \frac{571137}{430}  \, \zeta _3 - \frac{81108}{43}  \, \zeta _5 \,.
\end{align}
We further note that ${\cal N}_{m k}^{m^{\prime} k^{\prime}} = {\cal N}_{m^{\prime} k^{\prime}}^{m k}$,
${\cal R}_{m k}^{m^{\prime} k^{\prime}} = {\cal R}_{m^{\prime} k^{\prime}}^{m k}$,
 ${\cal \widetilde{N}}_{m k}^{m^{\prime} k^{\prime}} = {\cal \widetilde{N}}_{m^{\prime} k^{\prime}}^{m k}$,
and  ${\cal \widetilde{R}}_{m k}^{m^{\prime} k^{\prime}} = {\cal \widetilde{R}}_{m^{\prime} k^{\prime}}^{m k}$
due to the apparent charge-conjugation symmetry of the nucleon electromagnetic form factors.

\end{widetext}

\bibliographystyle{apsrev4-1}

\bibliography{References}

\end{document}